\documentclass[usenatbib]{mnras}

\usepackage{txfonts}
\usepackage[T1]{fontenc}
\usepackage{ae,aecompl}
\usepackage{graphicx}	
\usepackage{savesym}
\savesymbol{iint}
\savesymbol{iiint}
\savesymbol{iiiint}
\savesymbol{idotsint}
\usepackage{amsmath}	
\usepackage{amssymb}	
\usepackage{graphicx}
\usepackage{dcolumn}
\usepackage{bm}
\usepackage{subfigure}
\usepackage{float}
\usepackage{ulem}

\title[HCN Hyperfine Analysis of Massive Clumps]{HCN Hyperfine Ratio Analysis of Massive Molecular Clumps}

\author[Schap et al.]{\hspace{-4mm} \LARGE
W. J. Schap III,$^{1}$
P. J. Barnes,$^{1,2}$
A. Ordo\~{n}ez,$^{1}$
A. Ginsburg$^{3}$
Y. Yonekura $^{4}$
and Y. Fukui $^{5}$
\\
$^{1}$Astronomy Department, University of Florida, P.O. Box
112055, Gainesville, FL 32611, USA\\
$^{2}$School of Science and Technology, University of New England,
Armidale NSW 2351, Australia\\
$^{3}$European Southern Observatory, Karl-Schwarzschild-Strasse
2, D-85748 Garching bei M\"unchen, Germany\\
$^{4}$Center for Astronomy, Ibaraki University, 2-1-1 Bunkyo, Mito, Ibaraki 310-8512, Japan\\
$^{5}$Department of Astrophysics, Nagoya University, Furo-cho, Chikusa-ku, Nagoya 464-8602, Japan}

\date{\today}

\pubyear{2016}

\begin{document}
\label{firstpage}
\pagerange{\pageref{firstpage}--\pageref{lastpage}}
\maketitle

\begin{abstract}

We report a new analysis protocol for HCN hyperfine data, based on the \textsc{PySpecKit} package \citep{b7}, and results of using this new protocol to analyze a sample area of 7 massive molecular clumps from the CHaMP survey \citep{b1}, in order to derive maps of column density for this species. There is a strong correlation between the HCN integrated intensity, $I_{\rm HCN}$, and previously reported $I_{\rm HCO^{+}}$ in the clumps, but $I_{\rm N_{2}H^{+}}$ is not well-correlated with either of these other two ``dense gas tracers''. The four fitted parameters from \textsc{PySpecKit} in this region range over $V_{\rm LSR}$ = 8--10 km/s, $\sigma_V$ = 1.2--2.2 km/s, $T_{\rm ex}$ = 4--15 K, and $\tau$ = 0.2--2.5. These parameters allow us to derive a column density map of these clouds, without limiting assumptions about the excitation or opacity. A more traditional (linear) method of converting $I_{\rm HCN}$ to total mass column gives much lower clump masses than our results based on the hyperfine analysis. This is primarily due to areas in the sample region of low $I$, low $T_{\rm ex}$, and high $\tau$. We conclude that there may be more dense gas in these massive clumps not engaged in massive star formation than previously recognized. If this result holds for other clouds in the CHaMP sample, it would have dramatic consequences for the calibration of the Kennicutt-Schmidt star formation laws, including a large increase in the gas depletion timescale in such regions.

\end{abstract}




\begin{figure*}
\centering
\subfigure[]{
\includegraphics[width=8.4cm]{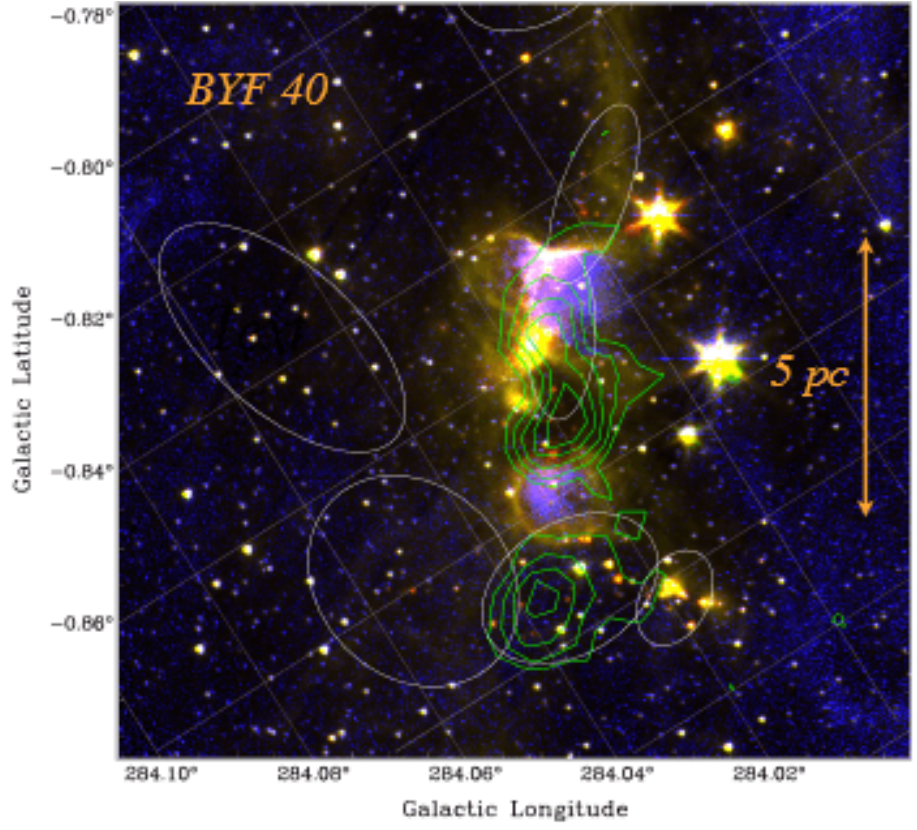}
\label{byf40}}
\subfigure[]{
\includegraphics[width=8.5cm]{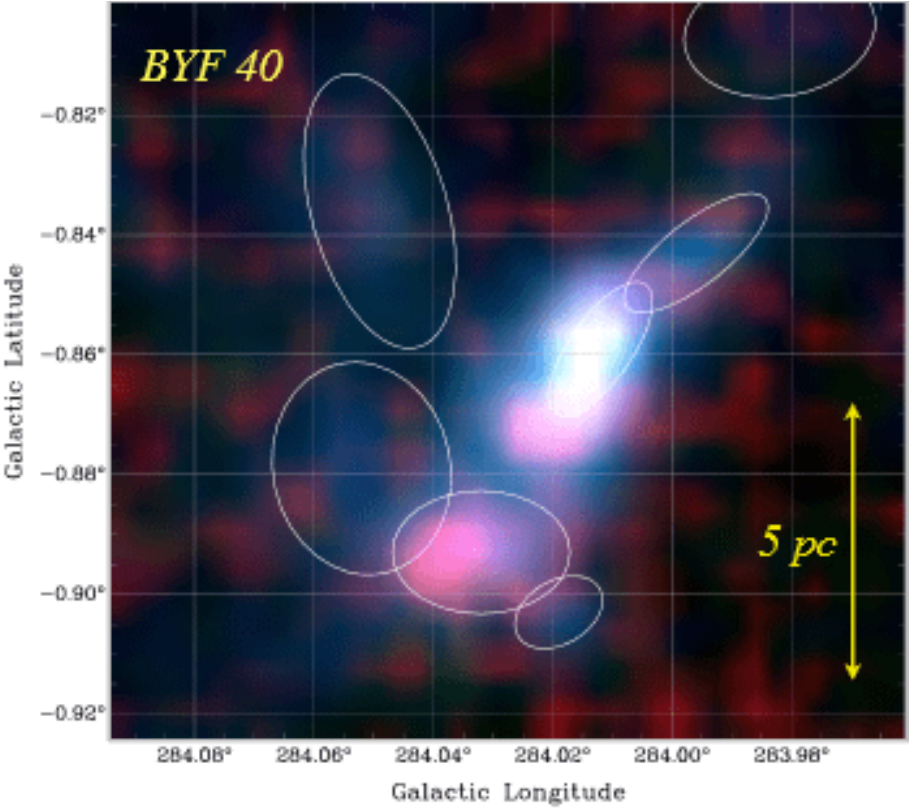}
\label{mopra}}
\caption{Three-color composite images of the BYF\,40 area. Figure (a) (red=IRAC2, green=IRAC1, blue=Br$\gamma$) includes N$_{2}$H$^{+}$  contours shown in green and HCO$^{+}$ clumps' emission HPW shown by ellipses (From Paper 1 and \citep{b2}). It has been transformed from RA/Dec in order to show Galactic coordinates, and the grid is tilted with respect to the image frame. \color{black} The scale bar is for an adopted distance of 6.6 kpc. Figure (b) presents a composite of Mopra-only maps of the seven BYF\,40 clumps a--g, where r=N$_{2}$H$^{+}$, g=HCN, and b=HCO$^{+}$ integrated intensities, along with the same HCO$^{+}$ HPW ellipses shown in (a). The HCO$^{+}$ and HCN emission track each other very closely (overall cyan coloring), while the N$_{2}$H$^{+}$ distribution is quite different. }
\label{3color}
\end{figure*} 

\section{\label{sec:level}Introduction}

In the distribution of molecular mass across the Milky Way Galaxy, dense molecular clumps are the sites where the gas begins its gravitational collapse and produces new stars.  Several studies have shown that these clumps not only create individual stars, but that many of these stars formed together as clusters \citep{b3, b5, b8}. It is important to understand the processes within and characteristics of clumps that form star clusters. This process of star formation is a driving force behind galactic evolution, and describes the global conversion of gas into stars described by the Kennicutt-Schmidt (KS) relations \citep{ke, b11}. 

Considering this significant effect on galactic evolution, it is important to understand the physical conditions and kinematics within these dense clumps in their prestellar phase. Until recently, unbiased studies of molecular emission at millimeter wavelengths have typically been for a small number of sources \citep{b16, b13, b9}, while others have chosen sources based on emission at other wavelengths \citep{b4, b17, b6, b14, b12, b19}, creating the possibility of statistical biases. Due partly to these shortcomings, there has been little consensus on the basic formation mechanism of stars.

Recently, however, this statistical limitation has begun to be addressed in several large-scale surveys of molecular line emission in the Galactic Plane \citep[e.g.,][]{j1,j2}. Among these, the Galactic Census of High- and Medium-mass Protostars \citep[CHaMP][hereafter Paper I]{b1} was specifically developed in order to better resolve such issues. With its large-scale, complete, unbiased, uniform, sensitive, and multi-wavelength coverage, CHaMP gives the properties of these dense molecular clumps and the stars that are produced from them. It also allows us to identify the evolutionary stages, and for the first time derive the time scale of the stages with a demographic analysis.

Some of the CHaMP data include molecular emission lines with hyperfine-split components, such as HCN and N$_{2}$H$^{+}$. By using hyperfine-split lines, spatially resolved maps of physical conditions and chemistry can be created.  These maps can be compared with HCO$^{+}$ and other species where simple rotational line analysis also yields abundances and kinematics (see Paper I).

From the CHaMP survey, BYF\,40 (part of Region 6, at Galactic coordinates near $l$ = 284.0, $b$ = --0.8) was selected for this initial analysis. While BYF\,40 is one of the more distant clouds ($\sim$6.6\,kpc) in the CHaMP sample, it is otherwise fairly typical of the brighter HCN-emitting clouds. Our objective was to measure and analyze the HCN column density of BYF\,40 in the survey, examine the science leading from these results, and to create usable code that would allow us to expand this analysis to rest of the survey.

\section{\label{sec:leve2}Observations and Data Reduction}

In Paper I we described the observational setup with the Mopra\footnote{The Mopra telescope is part of the Australia Telescope which is funded by
the Commonwealth of Australia for operation as a National Facility managed
by CSIRO. The University of New South Wales Digital Filter Bank used for
the observations with the Mopra telescope was provided with support from the
Australian Research Council.} radio telescope, including the configuration of the MOPS spectrometer, which collects data simultaneously on multiple transitions within an 8 GHz segment of the 3mm band. In Phase I of the observing (2004--2007), Mopra was used to collect data on several ``dense gas tracer'' spectral lines in the 85--93 GHz range, including HCN $J$=1$\rightarrow$0. Paper I fully describes the observing and data reduction procedures, including Mopra's on-the-fly (OTF) mapping capabilities, the Livedata/Gridzilla software pipeline, and other details. A brief summary of these procedures is outlined next.

BYF\,40 contains 7 subclumps, $a-g$, across an area of $\sim$0.2$^{\circ}$, with masses of a few $\times$ $10^3$ M$_\odot$ each (see Figure \ref{3color} and Table 1). The raw data were processed with the standard Livedata/Gridzilla pipeline. At 88.6 GHz the various telescope efficiencies are essentially the same as for the HCO$^+$ line at 89.2 GHz, as reported in Paper I. Four OTF fields of size $\sim5$ arcmin each took about 3 hours to complete with two orthogonal OTF mapping passes, and adjacent fields were composited in Gridzilla. The final product is a calibrated spectral line data cube (${l,b,V}$) converted to the $T_r^*$ scale, with a mass of 2000-7000 M$_\odot$ per clump (based on the HCO$^{+}$ analysis). Normally, such data cubes would then be amenable to moment analysis, but the hyperfine lines contain additional information and artificially widen the otherwise gaussian profile. We describe our alternative hyperfine approach next. 

The HCN $J$=1$\rightarrow$0 line is not a simple transition, since the $^{14}$N nucleus has a large nuclear quadrupole moment which induces a hyperfine splitting of the rotational levels. Thus, species such as HCN and N$_{2}$H$^{+}$ have hyperfine structure which is not well served by the methods used in Paper I. In this case we have 3 hyperfine components, which will have intrinsic line strengths in the optically thin limit of 1:5:3, assuming they are formed in local thermodynamic equilibrium (LTE).  Radiative transfer analysis of all three components allows solutions for both the lines' excitation temperature ($T_{\rm ex}$) and optical depth ($\tau$), but such treatments exist in only a few software packages, not all of which are easy to use.

\begin{figure*}
\hspace{-100mm}
\subfigure[]{
\includegraphics[width=7.2cm, angle=-90]{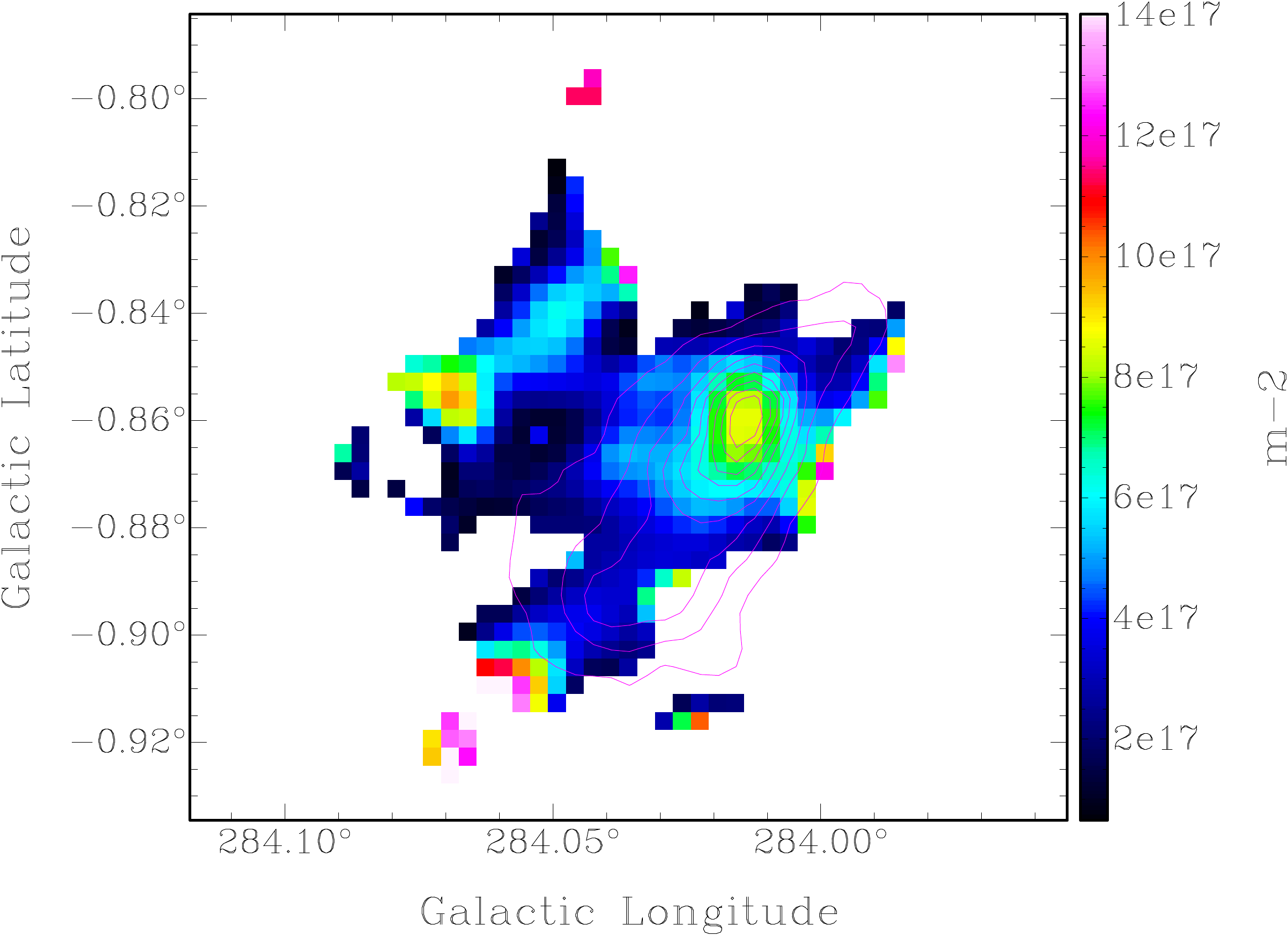}
\label{ncolcon}}
\subfigure[]{
\includegraphics[width=7.2cm, angle=-90]{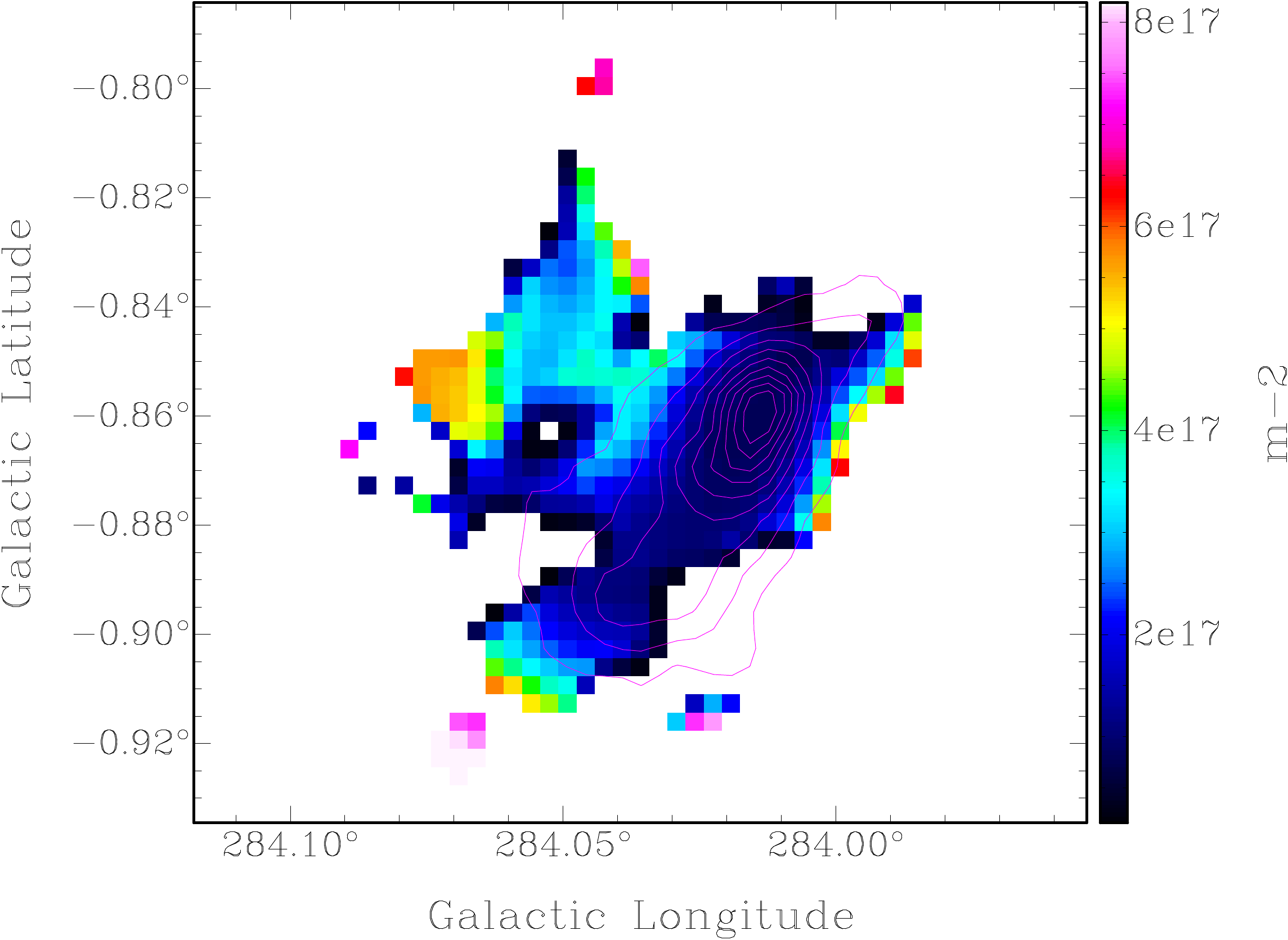}
\label{encol}}
\hspace{-100mm}
\caption{Maps of $N_{\rm HCN}$ (a) and its uncertainty (b) across BYF\,40, plotted with contours at intervals of 10\% of the peak $I_{\rm HCN}$.}
\label{ncol}
\end{figure*}

\section{\label{sec:leve3}Analysis Methods}

\subsection{\label{sec:leve4} \rm \textsc{PySpecKit} Analysis}

The \textsc{PySpecKit} package \citep{b7} was used for the hyperfine line fitting of the BYF\,40 HCN $({l,b,V})$ data cube we chose as a test area for our code development, and has the advantage of being scriptable in a modern computing environment, as well as having a built-in tool for fitting the model across all spectra in a cube. Assuming LTE, it solves for the common parameters of all hyperfine components.

\begin{figure}
\centering
\hspace*{-0.75cm}
\includegraphics[width=10cm]{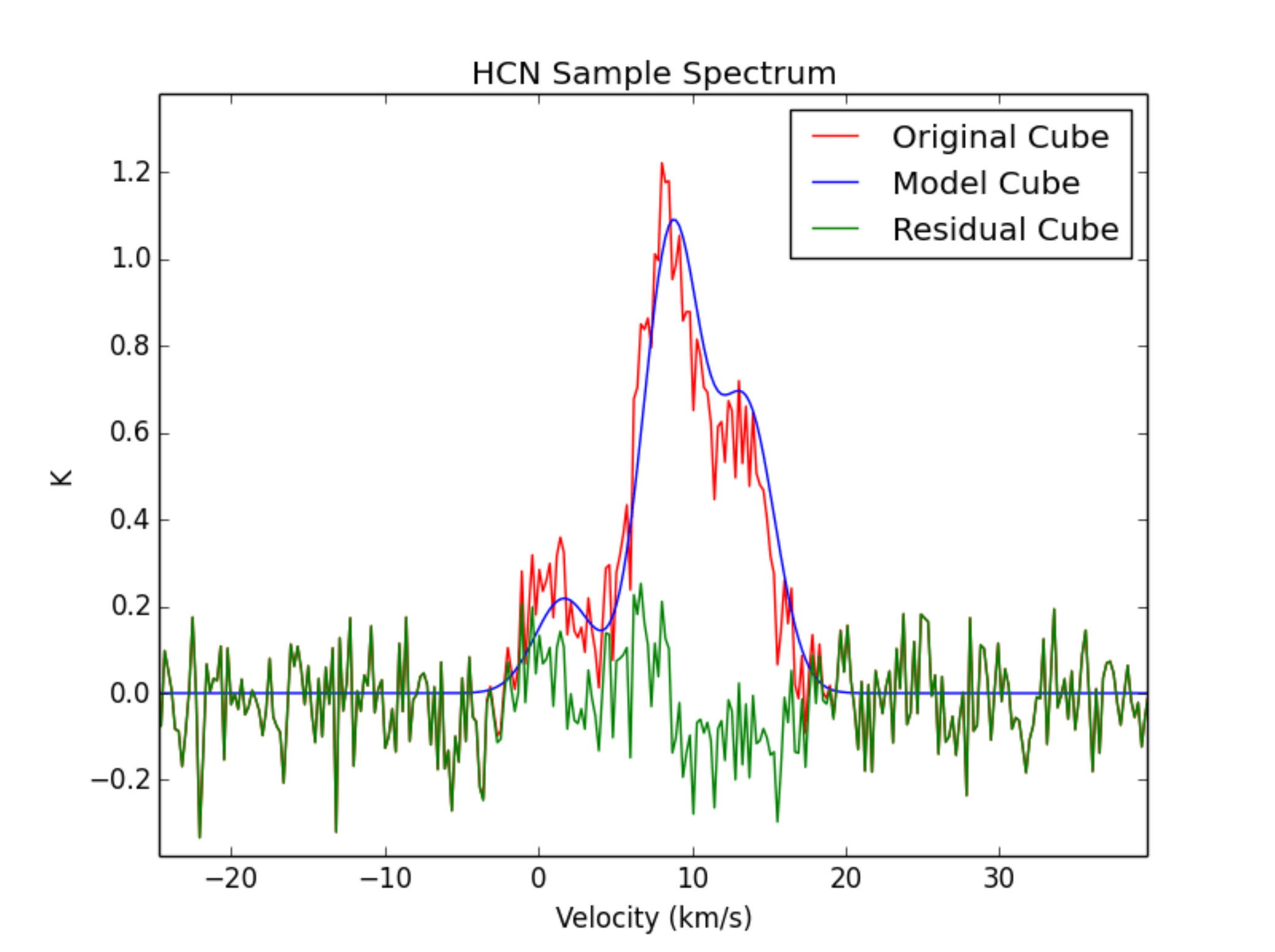}
\vspace{-5mm}
\caption{Sample spectra from the original, model, and residual data cubes. Each was measured at the coordinate $l$=284.002(3),$b$=--0.846(3).}
\label{spectra}
\end{figure}

In order to improve the fit quality, the original FITS cube input for the package was binned by two spectral channels and convolved from an original resolution of 40$''$ to 60$''$. Once a minimum S/N threshold is set for the hyperfine fitting, \textsc{PySpecKit} yields separate 2D maps of the excitation temperature $T_{\rm ex}$, opacity $\tau$, mean velocity $V_{\rm LSR}$, and velocity dispersion $\sigma$$_{V}$, as well as error maps for each parameter map. A sample fit is shown in Figure \ref{spectra}, while the parameter maps appear in Figures \ref{vlsr}--\ref{tex}.

\subsection{\label{sec:leve5} \rm Column Density Calculation}

After obtaining the parameter maps and uncertainties as outputs from \textsc{PySpecKit}, the column density $N$ was evaluated using a mix of c-shell and \textsc{Miriad} \citep{b15} software. $\tau$ and $T_{\rm ex}$ were measured in each voxel and then converted into $N_{\rm HCN}$ using the equation

\begin{equation}
N_{\rm HCN} = 9.0\times10^{15}\,{\rm m}^{-2}~\frac{Q(T_{\rm ex})e^{E_{\rm up}/kT_{\rm ex}}}{1-e^{-h\nu/kT_{\rm ex}}}~ \int \tau dV
\end{equation}

Here, $Q$ refers to the partition function, $E_{\rm up}$ is the energy of the upper level above the ground state of the $J_{\rm upper}$$\rightarrow$$J_{\rm lower}$ transition, $h\nu$ is the energy of the transition, and $k$ is the Boltzmann constant. The derived $N_{\rm HCN}$ map is shown in Figure \ref{ncol}, and has a flatter distribution compared to the overlayed $I_{\rm HCN}$ contours.

Under traditional methods, the $X$ factor for $^{12}$CO is often used to estimate column densities with a simple linear conversion. The accuracy of such conversions have frequently been debated, and recent work by our group \citep{bm1,bh1} has shown that, in the case of CO at least, both the lines' $T_{\rm ex}$ and $\tau$ need to be properly accounted for, and must be self-consistently measured.

In this paper, we calculate the uncertainty in $N_{\rm HCN}$ by propagating all of the formal fit errors, and applying the constraints $T_{\rm ex}$ \textgreater 2.73 K and $\tau$ \textgreater 0 on the \textsc{PySpecKit} solutions. By using the error images and the output images together, relative errors were also determined for several intermediate quantities in the calculation, and for $N_{\rm HCN}$. The  $V_{\rm LSR}$ values range from 8--10 km/s, $\sigma_V$ = 1.2--2.2 km/s, $T_{\rm ex}$ = 4--9K, and $\tau$ = 0.5--2.5.

\subsection{\label{sec:leve6} \rm Clump Masses}

\begin{figure*}
\centering
\subfigure[]{
\includegraphics[width=8.5cm]{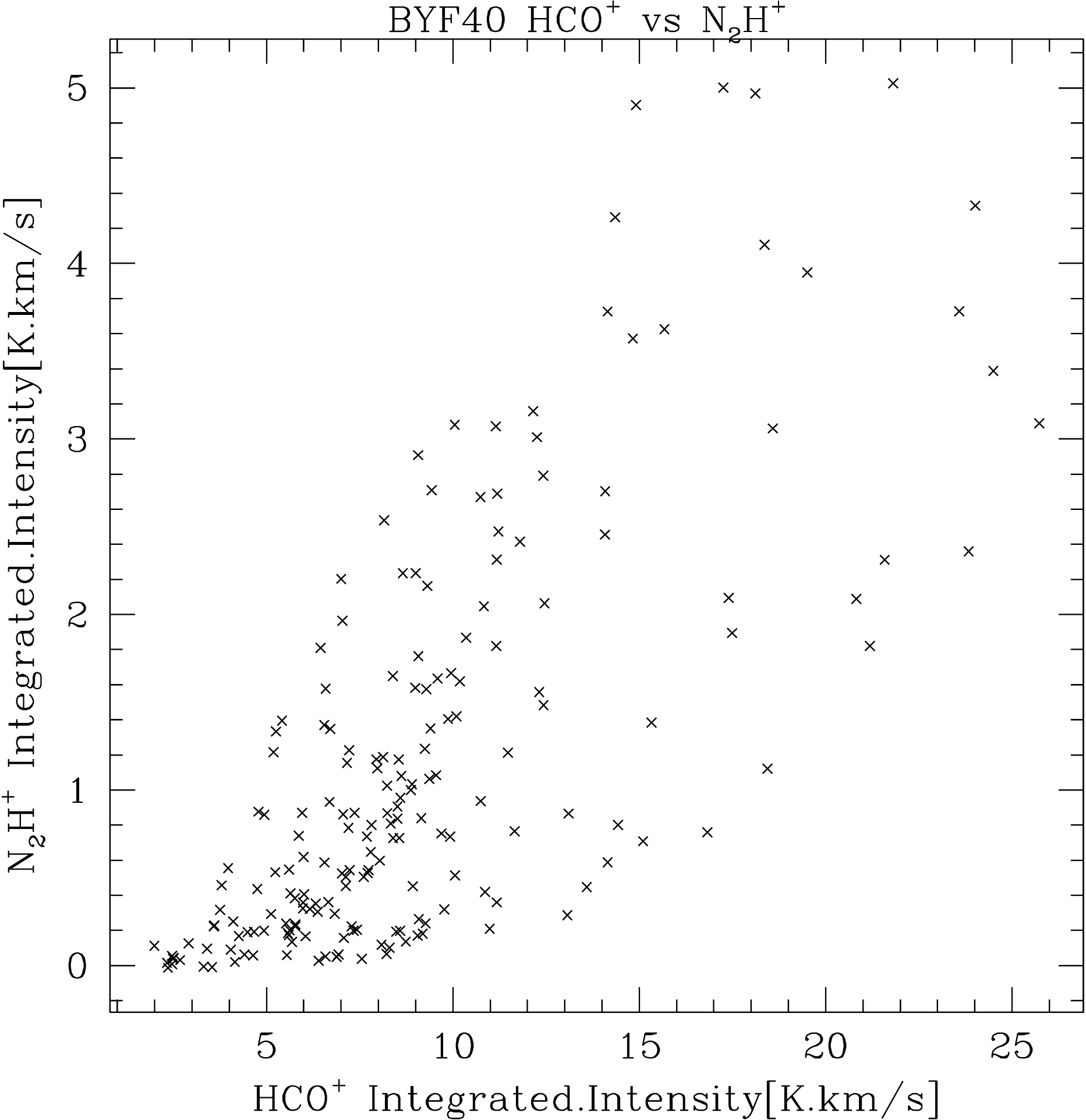}
\label{hcovsn2hp}}
\subfigure[]{
\includegraphics[width=8.5cm]{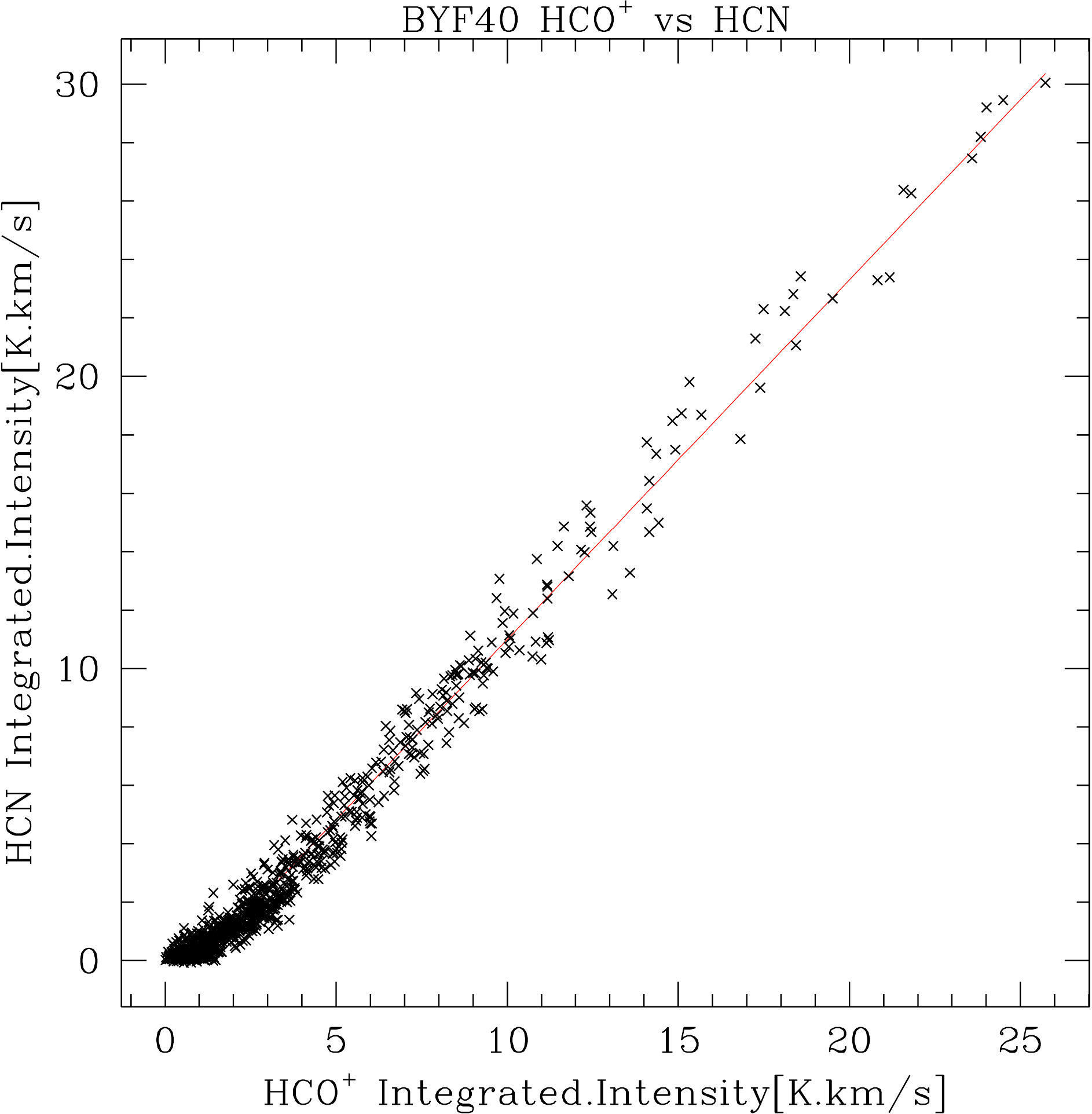}
\label{hcovshcn}}
\vspace{-1mm}
\caption{The pixel by pixel integrated intensity of HCO$^{+}$ vs N$_{2}$H$^{+}$ and HCO$^{+}$ vs HCN across the BYF\,40 area. The values from Figure 4(a) have a correlation coefficient r$^{2}$ of 0.52. Figure 4(b) however, includes a weighted least squares linear fit (shown as a red line) of slope 1.232$\pm$0.006 and intercept --1.33$\pm$0.04 K km/s, and r$^{2}$ = 0.98.}
\label{hcopvs}
\end{figure*}

 In Table 1 we give the observed and physical parameters of the BYF\,40 clumps used to derive these results. The values appear similar to Table 4 from Paper I, due to the high correlation present between HCN and HCO$^{+}$ in the clump. Additionally, Figure \ref{hcovshcn} shows that there is a high correlation between the integrated intensities of HCN and HCO$^{+}$. This can be seen in the  \ref{byf40} Mopra map, where the blue of HCO$^{+}$ occurs frequently in proportion to the green HCN, yielding the overall cyan color in this rendering.

Integrating the column density map across each clump gives the total clump mass, and a total mass column measurement across the map ($M_{\rm map}$ = 57,000 M$_ \odot$), assuming an intrinsic HCN abundance of 10$^{-9}$, the same as for HCO$^{+}$. This result is roughly twice as large as the 24,000-37,000 M$_ \odot$ value obtained when converting the $I_{\rm HCN}$ using a standard linear conversion. This is caused by significant areas in Region 6 of low $I$ and low $T_{\rm ex}$, but very high $\tau$. It should be noted that this is an overall result, where some measured clumps are more or less massive than the value linearly converted from $I$. This shows that when simple $I$ scaling is used to estimate the mass in a clump, the mass can be substantially underestimated, and suggests that there is more dense gas in the clumps not engaging in massive star formation than previously measured.

\begin{table*}
 \caption{BYF40 Observed and Derived HCN Parameters}
 \label{tab:journal_abbr}
 \begin{tabular}{cccccccccccc}
    \hline
     Clump & l    & b      & I          & T$_{\rm peak}$ & Radius$^a$   & V$_{\rm LSR}$  & $\sigma$$_{\rm V}$  & $\tau^b$    & T$_{\rm ex}^b$  & N$_{\rm HCN}^b$   & Mass\\
        & (deg)   & (deg)  & ($K*km s^{-1}$) & (K)        & (pc)         & (km $s^{-1}$)    & (km $s^{-1}$)       &             & (K)             & ($m^{-2}$)        & (10$^3$ M$_ \odot$)\\
    \hline
a  & 284.012(3) & -0.859(3) & 30.05(43)    & 3.722(30)  &  1.09(19)      & 9.02( 4)       & 2.10( 4)          & 1.22(20)    & 7.1(5)          & 8.6E+17(8)      & 34.9(56)  \\
b  & 284.032(3) & -0.893(3) & 11.07(35)    & 2.04(30)   &  1.41(19)      & 9.25(30)       & 1.66( 5)          & 0.60(42)    & 7.2(27)         & 6.9E+17(3)      & 46.9(52)  \\
c  & 284.019(3) & -0.903(3) &  4.27(24)    & 1.26(30)   &  0.68(19)      & 9.14( 6)       & 1.41( 7)          & --          & --              & --              & --  \\
d  & 283.996(3) & -0.839(3) &  6.57(33)    & 1.73(32)   &  0.81(19)      & 8.26( 9)       & 1.87( 9)          & 0.42(27)    & 6.8(45)         & 2.2E+17(3)      & 5.0(11)  \\
e  & 284.056(3) & -0.876(3) &  3.69(24)    & 0.99(26)   &  1.30(19)      & 9.60( 8)       & 1.67( 5)          & 0.94(58)    & 5.2(26)         & 2.1E+17(2)      & 10.1(15)  \\
f  & 284.049(3) & -0.833(3) &  2.57(19)    & 1.08(26)   &  1.10(19)      & 8.29(10)       & 1.42( 9)          & 1.97(85)    & 3.6(3)          & 6.1E+17(28)     & 25(12)  \\
g  & 283.986(3) & -0.816(3) &  1.01(14)    & 0.97(32)   &  1.04(19)      & 9.63(14)       & 1.37(15)          & --          & --              & --              & --  \\
    \hline
       \multicolumn{12}{c}{$^a$ Radii computed from angular size and a distance of 6.6 kpc.}\\
       \multicolumn{12}{c}{ $^b$ Values for clump $b$ ($T_{\rm ex}$, $\tau$), $d$ ($T_{\rm ex}$,  $\tau$, $N_{\rm HCN}$), and $e$ ($\tau$, $T_{\rm ex}$) are taken from an offset position.}\\  
 \end{tabular}
\end{table*}

\begin{figure} 
\centering
\includegraphics[width=9cm]{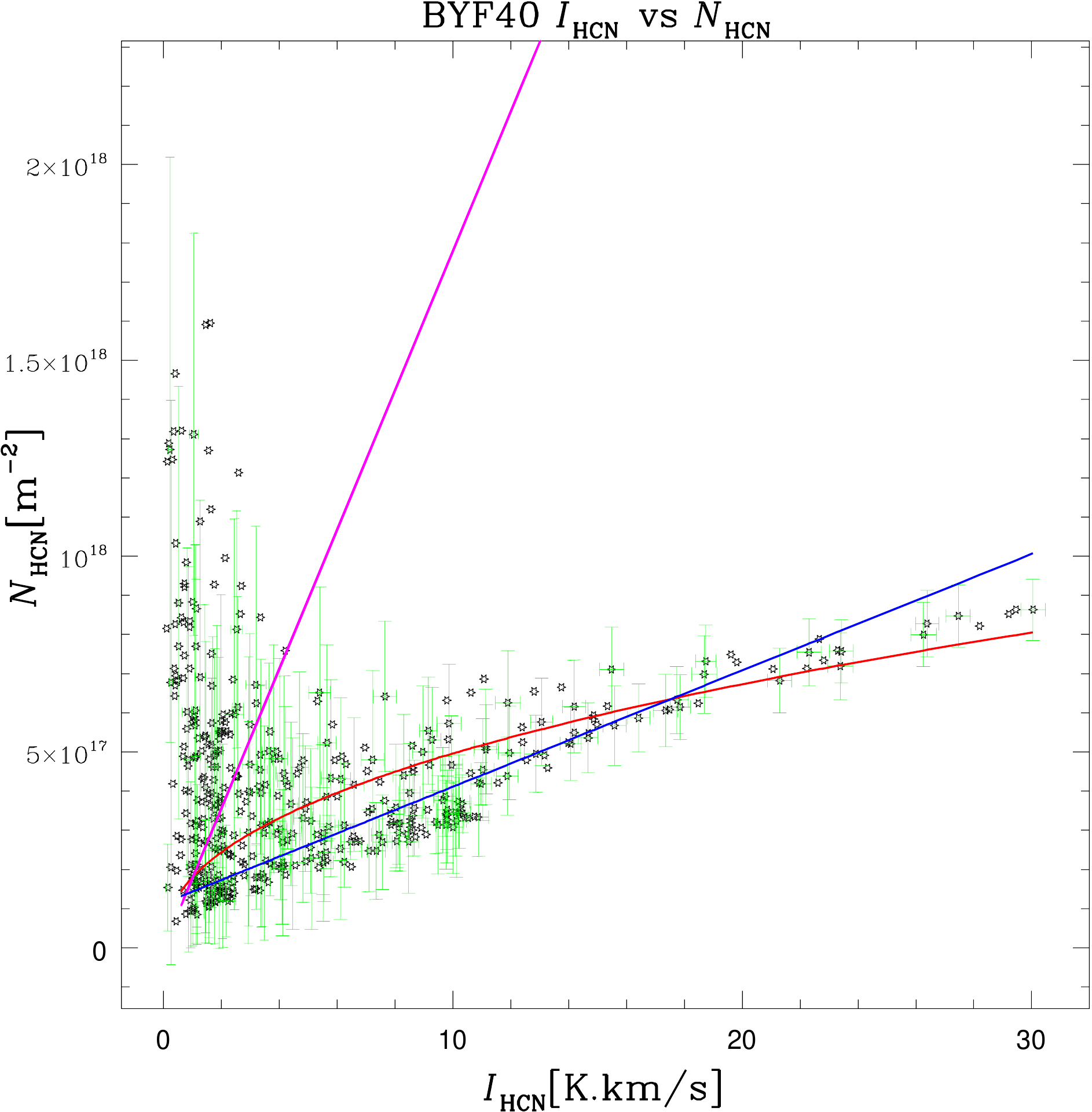}
\vspace{-5mm}
\caption{Pixel by pixel integrated intensity vs column density from the HCN hyperfine results which shows a weighted least squares linear fit (blue) with slope = (2.97$\pm$0.08)$\times$10$^{16}$ molecules m$^{-2}$ (K km/s)$^{-1}$, intercept = (1.14$\pm$0.06)$\times$10$^{17}$ molecules m$^{-2}$, and r$^{2}$ = 0.10 ; a linear zero-intercept fit (magenta) with slope = (1.8$\pm$2.5) $\times$10$^{17}$ molecules m$^{-2}$ (K km/s)$^{-1}$; and a power law fit (red) of the form log[$N$/(molecules m$^{-2}$)] = (0.44$\pm$0.02)*log[$I$/(K km/s)]+17.25$\pm$0.02 and r$^{2}$ = 0.01.}
\label{3fit}
\end{figure}

\section{\label{sec:leve7}Discussion}

\subsection{\label{sec:leve8} \rm Column density vs. opacity and excitation}

It is often challenging to evaluate equation (1), since one needs ways to separately estimate $\tau$ and $T_{\rm ex}$. In the literature it is often assumed, in the radiative transfer equation
\begin{equation}
kT_b/h\nu=(J_{\rm source}-J_{\rm bkgd})(1-e^{-\tau)},
\end{equation}
that $\tau$ $\ll$ 1, so that $\tau$ can be approximated by $T_b$/$T_{\rm ex}$.  Then with the excitation-dependent terms in equation (1) approximated as being linear in $T_{\rm ex}$ (for reasonably high T$_{\rm ex}$), $N$ can be obtained via
\begin{equation}
N \propto \int T_{\rm b}dV,
\end{equation}
so that $\tau$ and $T_{\rm ex}$ are not explicitly required, unless the emission is in the low-$T_{\rm ex}$ and/or high-$\tau$ regime. However, if we take advantage of hyperfine-line splitting and assume the components are formed in LTE, $\tau$ and $T_{\rm ex}$ can be obtained directly, along with $V_{\rm LSR}$ and $\sigma_V$, allowing a more direct measure of $N$, and mass estimates to be obtained from relative abundance maps or vice versa.

Figure \ref{3fit} shows that when calculating the mass estimate, relying on a simple $I$ scaling may substantially underestimate the mass of gas in these clumps, especially where $I$ is low (but still at reasonably high S/N), $T_{\rm ex}$ is low, but $\tau$ is high. Figure \ref{3fit} therefore shows exactly why one needs to consider the low-$T_{\rm ex}$ or high-$\tau$ case when evaluating $N$ in cold molecular clouds.  It also shows that the uncertainty measure of $N$ from our method is low enough that the numerous high-$\tau$ pixels are not noise-dominated. Thus, there is a significant amount of dense gas in these massive clumps that is not obviously engaged in massive star formation, as traced (e.g.) by Br$\gamma$ emission or bright (high-$T_{\rm ex}$) molecular lines.

These hyperfine results allow us to compare the column densities and kinematics of HCN as presented here with the HCO$^{+}$ from Paper I, with the N$_{2}$H$^{+}$ from \citet{b2}, and with other unpublished data on Region 6/BYF\,40. Several different molecular tracers in these clouds, such as HCN, $^{13}$CO, and HCO$^{+}$, seem to have a similar $J$=1$\rightarrow$0 morphology and distribution, as demonstrated in Figures \ref{mopra} and \ref{hcovshcn}, while \citet{b23} shows similarities between HCN $J$=3$\rightarrow$2 and HCO$^{+}$ $J$=3$\rightarrow$2. However, as can be seen in Figures \ref{3color} and \ref{hcovsn2hp}, the N$_{2}$H$^{+}$ $J$=1$\rightarrow$0 has a very different distribution. This disparity among N$_{2}$H$^{+}$ and other molecular species is similar to the results of \citet{wo}, who were one of the first to find chemical discrepancies between HCO$^{+}$ and N$_{2}$H$^{+}$; it has also been reported in other contexts, such as \citet{b23}.

Since HCO+ and HCN are very tightly correlated in BYF 40, the correlation between HCO+ and Br$\gamma$ seen in \citet{b2} must now extend to HCN and $^{13}$CO as well.  This demonstrates that the ratio of N$_{2}$H$^{+}$ emission to that of the other species (including HCN) must be temperature- and/or ionisation-sensitive, while a similar chemical origin is shared between HCN, $^{13}$CO, and HCO$^{+}$.  Indeed, in contrast to HCO$^{+}$ and HCN, N$_{2}$H$^{+}$ tends to trace cooler, quiescent gas and must therefore express dissimilar chemical behaviour to the warm, energetic gas traced by HCO$^{+}$, HCN, and $^{13}$CO that is present within BYF 40. 

In the case of $^{13}$CO, this ionisation-sensitivity may not be so surprising. \citet{b2} noted that HCO$^{+}$, conventionally thought of as a cold and dense gas tracer, is unexpectedly well-correlated with MYSO photospheric tracers (i.e., Br$\gamma$). We now see that HCN shows the same relationship with, which is perhaps even more surprising given that it is a neutral species. HCN is assumed to be a cornerstone of the dense-gas Kennicutt-Schmidt (KS) star formation relations \citep[e.g.,][ and references therein]{b21, b22}, and $I_{\rm HCN}$ should be a better tracer of the dense gas column most directly engaged in active star formation than are lower density tracers like CO or HI. Therefore, the idea that HCN is measuring the same thing as Br$\gamma$ (a presumably direct tracer of star formation activity), calls into question the entire basis of the HCN version of the KS relations.

\subsection{\label{sec:leve9} \rm Consequences for the Kennicutt-Schmidt relation} 

We have used the HCN analysis to show that $I$ $\propto$ $N$ is not true, especially at small $I$.  This is very important for the rest of the argument about the KS relation, which says that $L_{\rm SFR}$ $\propto$ $\Sigma_{\rm SFR}^{p}$, where $p$ is some power.

We showed that in BYF\,40, $I_{\rm HCN}$ = const * $I_{\rm HCO^{+}}$.  But we showed in \citet{b2} that $I_{\rm HCO^{+}}$ = const * $I_{\rm Br\gamma}^{0.24\pm0.04}$. This is not as tight as the HCN/HCO$^{+}$ connection, but it is significant. Br$\gamma$ comes primarily from HII regions, which are ionized only by massive young stars, providing a relatively direct link between the observable line luminosity and the recent star formation rate. We therefore show that $I_{\rm HCN}$ is proportional (in the log) to SFR as traced by (e.g.) $L_{\rm Br\gamma}$ or $L_{\rm IR}$. In summary, the Br$\gamma$-$I_{\rm HCN}$ correlation suggests that HCN is a star-formation-driven feedback indicator.

Now consider the masses. HCN is the canonical ``dense gas tracer,'' so by convention, it is widely assumed to trace the mass involved in SF more directly than CO. \citep[E.g. see][ and references therein.]{ke}  This was also the justification for using $I_{\rm HCN}$ in the original {linear} \citet{gs} $L_{\rm IR}$-$I_{\rm HCN}$ version of the KS law in external galaxies. The canonical assumption is therefore: $I_{\rm HCN}$ $\propto$ $\Sigma_{\rm SF}$.

We contend that the \citet{gs} interpretation of the KS law is invalid. The linear correlation between $I_{\rm HCN}$ and $L_{\rm SFR}$ does not link star formation and dense gas mass.  Instead, the HCN luminosity is driven directly by stellar feedback and is therefore a direct tracer of star formation.\color{black}

Therefore, if HCN traces $L_{\rm SF}$ instead of $M_{\rm SF}$, the \citet{gs} result fails to establish a ``dense gas'' KS law and instead is a result of HCN excitation in the presence of SF.  More generally, and especially at low $I$, HCN strongly does $not$ trace mass (and is not even fitted by a power law).  N$_{2}$H$^{+}$'s poor correlation with HCO$^{+}$ shown in Figure \ref{hcovsn2hp} supports our argument that HCO$^{+}$ and HCN do not trace dense gas. To establish what the ``dense gas'' KS law really looks like, we must use a different method than just measuring $I_{\rm HCN}$ to find $\Sigma_{\rm SF}$, although instead it may be possible to use $I_{\rm HCN}$ or $I_{\rm HCO^{+}}$ to measure $L_{\rm SF}$.

Despite these arguments, our sample size is quite small, and our conclusions are tentative. However, once we apply the same analysis to the rest of the champ HCN data (Schap et al., in prep), we will have a much stronger statistical base over which to examine these arguments.

In the context of our column density results, we can conjecture that, should the existence of large masses of cold, non-starforming gas be confirmed, the normalisation of the KS relations could be shifted strongly towards higher mass/lower SFR in the $\Sigma$(gas)-$\Sigma$(SFR) plane. This would consequently lengthen the gas depletion timescale by a correspondingly large factor.

The next step will be to apply the methods outlined in this paper to HCN maps for the rest of the 303 CHaMP clumps.  Once the analysis of the HCN is complete it can be extended to other species, such as the N$_{2}$H$^{+}$ data. From there, column density and relative abundance maps can be calculated using these results, giving new mass estimates for these massive clumps, and will give a strong statistical basis for testing the physical underpinnings of the KS relations.
	
\vspace{-5mm}
\section{\label{sec:level0}Conclusion}

Using the \textsc{PySpecKit} package, we developed a new pipeline for HCN hyperfine line data analysis, and tested it on one of the regions of the CHaMP project, containing the 7 clumps that are part of the source BYF\,40. The intent is that this pipeline can next be used to analyse the rest of the regions of the survey. Following the full evaluation of HCN, the analysis can be extended to the CHaMP N$_{2}$H$^{+}$ data and more. 

We also examine the science derived from the HCN analysis, where the hyperfine fitting simultaneously and directly gives solutions for $V_{\rm LSR}$, $\sigma_V$, $T_{\rm ex}$, and $\tau$ at each pixel. Using these results, $N_{\rm HCN}$ can be directly measured across the mapped area, without having to convert from $I_{\rm HCN}$ and make assumptions about $\tau$ or $T_{\rm ex}$. 

Assuming an HCN abundance, we also compute the 7 clump masses. The total mass of BYF40 is calculated to be $\sim$57,000 M$_\odot$, roughly twice as large as the 24,000-37,000 M$_ \odot$ value obtained from converting the $I_{\rm HCN}$ in the traditional manner, due largely to the contribution to the total from areas with low $T_{\rm ex}$, but very high $\tau$, and hence high $N$. Thus, there is a significant amount of dense gas in the clumps not engaging in massive star formation, as traced by measures of excitation from YSOs. 

In addition, we find a very strong correlation between the integrated intensities of HCN and HCO$^{+}$, which has previously been shown to be correlated with Br$\gamma$ emission.  
This suggests that even $I_{\rm HCN}$ may not be an unbiased tracer of high-density, star-forming gas in the sense required by the Kennicutt-Schmidt relations. Our results suggest a recalibration of the KS relations may be necessary, implying also a longer gas depletion timescale in Galactic star forming regions. This and future results will be made publicly available for modellers on the CHaMP website, http://www.astro.ufl.edu/champ, providing the FITS files of data, derived quantities, and machine readable tables and catalogues.

\vspace{-5mm}
\section{\label{sec:level11} Acknowledgements}
PJB acknowledges support from NSF grant AST-1312597, and NASA-ADAP grant NNX15AF64G. In addition, we thank the anonymous reviewer for helpful suggestions and comments made on the paper.

\appendix 
\section{\label{sec:level2}}

\begin{figure*}
\centering
\subfigure[V$_{\rm LSR}$ Map]{
\includegraphics[width=6.5cm, angle=-90]{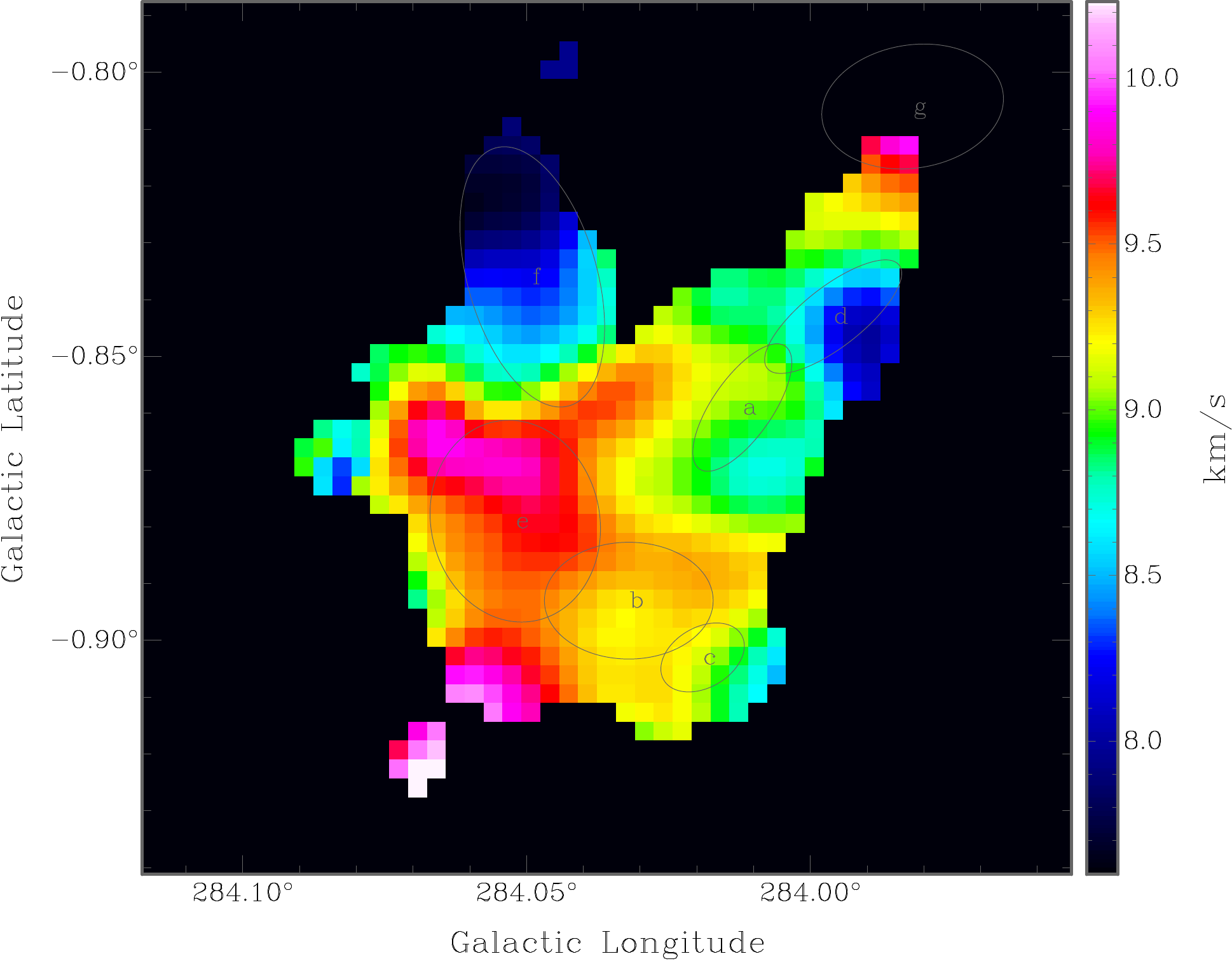}
\label{fig:subfigure1}}
\subfigure[V$_{\rm LSR}$ Uncertainty]{
\includegraphics[width=6.5cm, angle=-90]{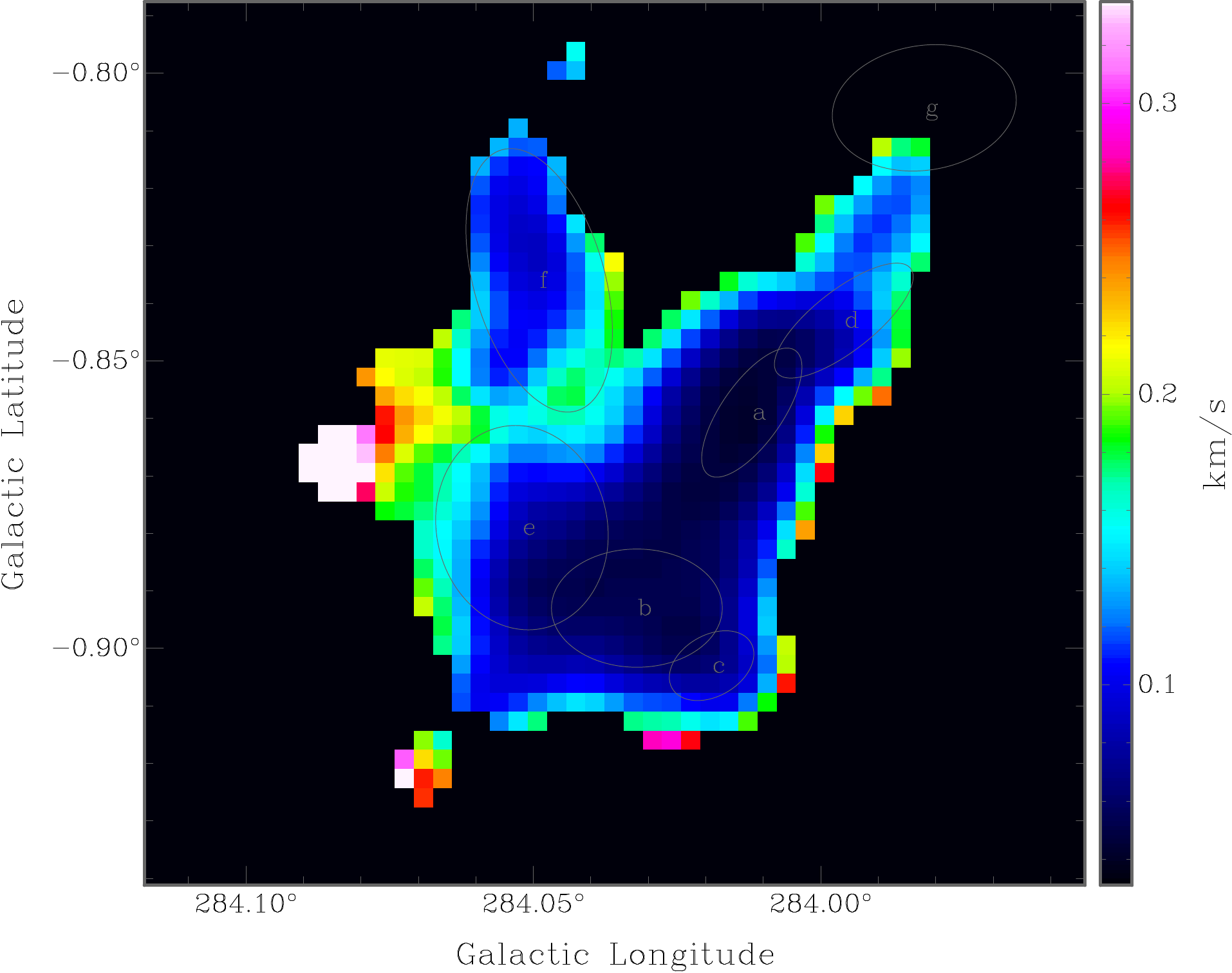}
\label{fig:subfigure2}}
\caption{BYF\,40 maps of V$_{\rm LSR}$ and its uncertainty given by \textsc{PySpecKit}, overlaid by the HCO$^{+}$ ellipses from Paper I.}
\label{vlsr}
\end{figure*} 
 
\begin{figure*}
\centering
\subfigure[Velocity Dispersion Map]{
\includegraphics[width=6.5cm, angle=-90]{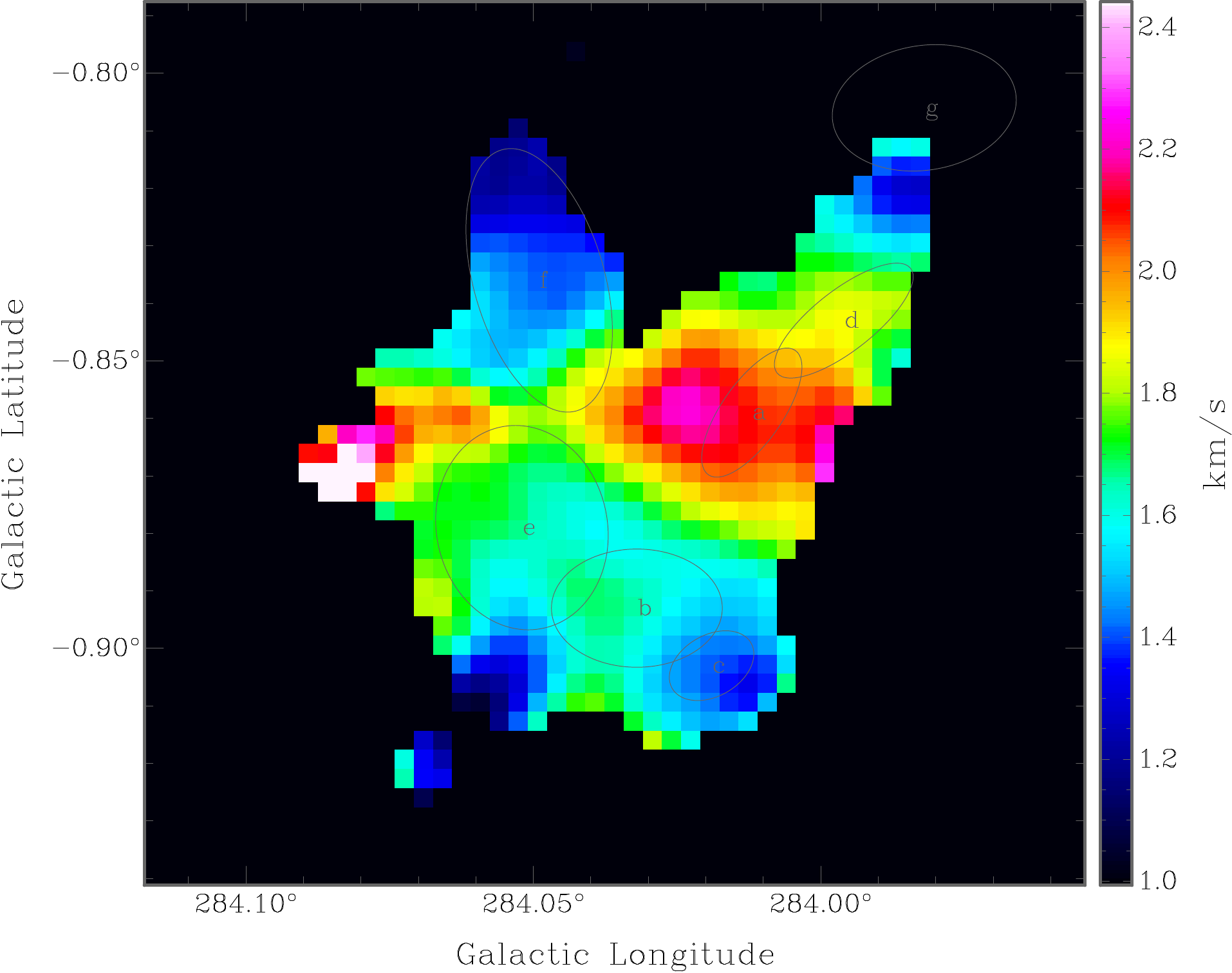}
\label{fig:subfigure3}}
\subfigure[Velocity Dispersion Uncertainty]{
\includegraphics[width=6.5cm, angle=-90]{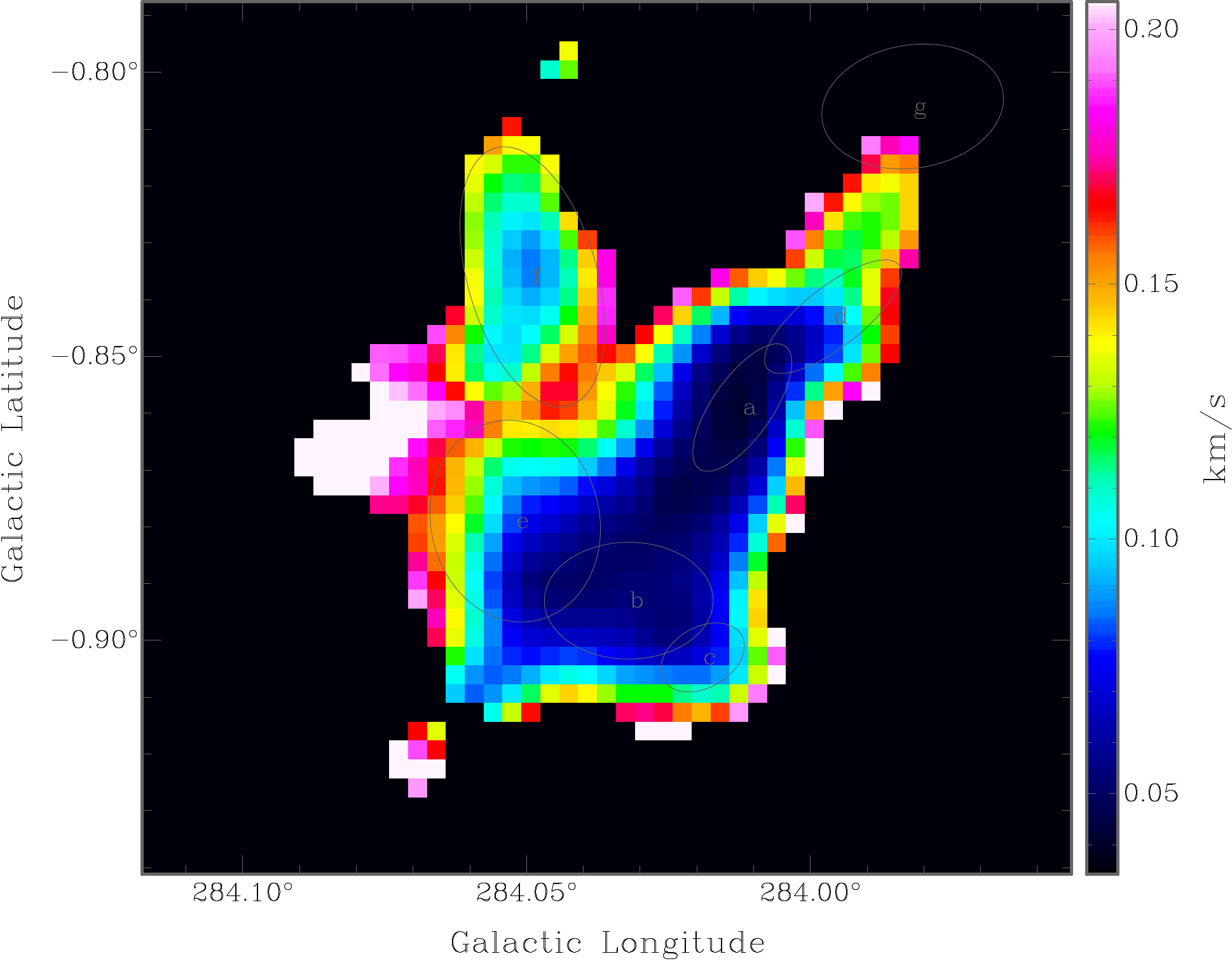}
\label{fig:subfigure4}}
\caption{BYF\,40 maps of the velocity dispersion $\sigma_V$ and its uncertainty given by \textsc{PySpecKit}, overlaid by the HCO$^{+}$ ellipses from Paper I.}
\label{dispersion}
\end{figure*}

\begin{figure*}
\centering
\subfigure[Opacity Map]{
\includegraphics[width=6.5cm, angle=-90]{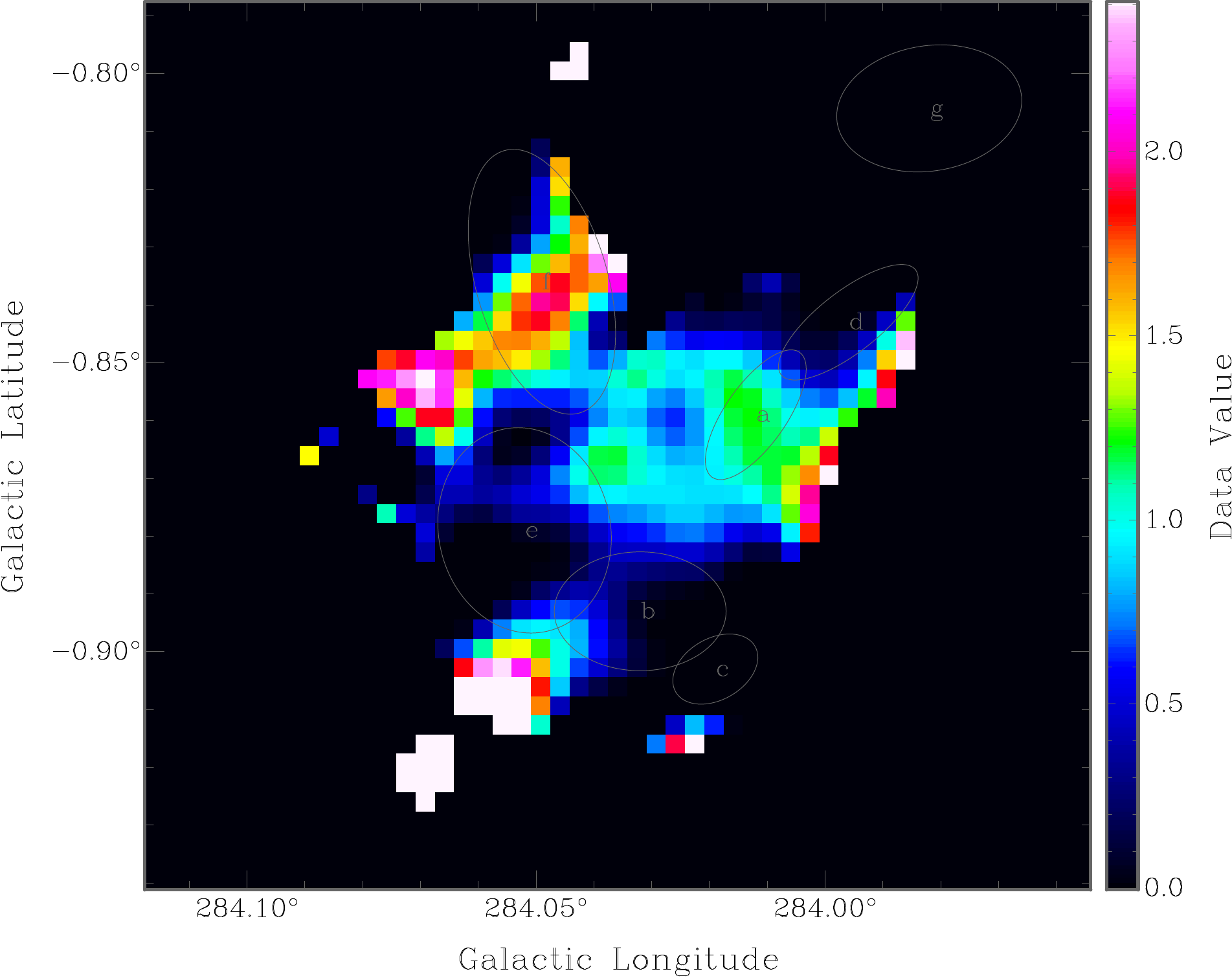}
\label{fig:subfigure5}}
\subfigure[Opacity Uncertainty]{
\includegraphics[width=6.5cm, angle=-90]{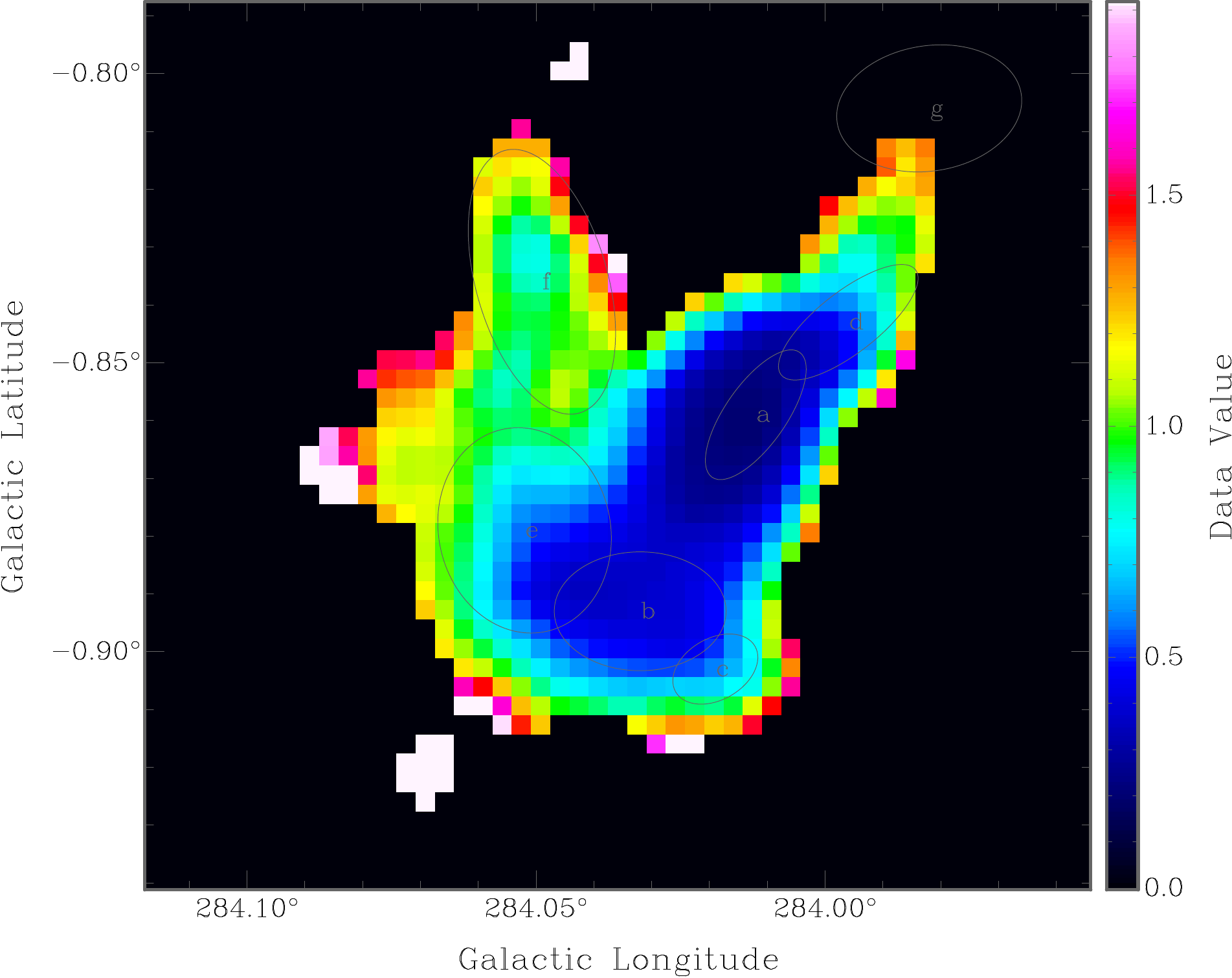}
\label{fig:subfigure6}}
\caption{BYF\,40 maps of opacity $\tau$ and its uncertainty given by \textsc{PySpecKit}, overlaid by the HCO$^{+}$ ellipses from Paper I.}
\label{opacity}
\end{figure*}

\begin{figure*}
\centering
\subfigure[T$_{\rm ex}$ Map]{
\includegraphics[width=6.5cm, angle=-90]{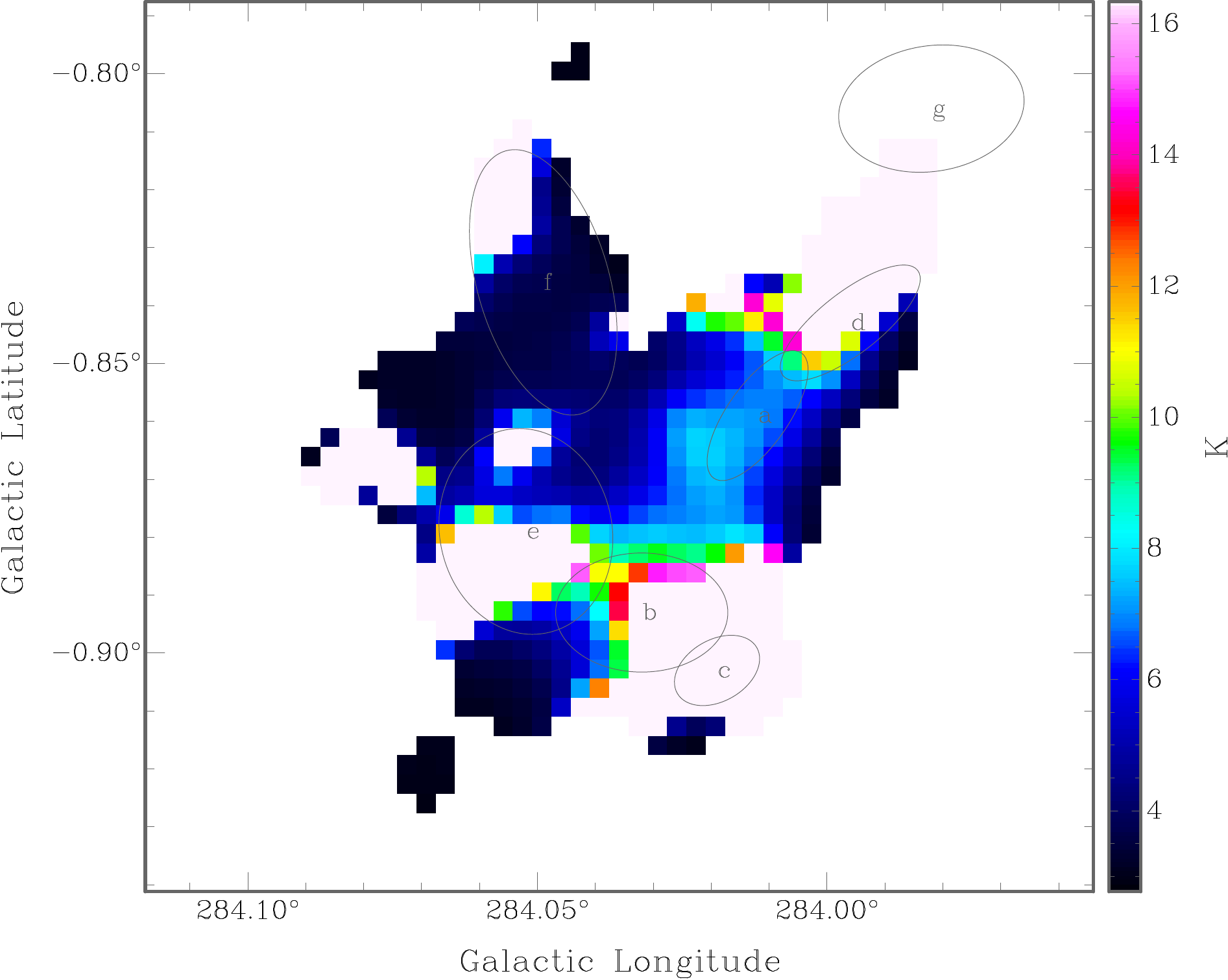}
\label{fig:subfigure7}}
\subfigure[T$_{\rm ex}$ Uncertainty]{
\includegraphics[width=6.5cm, angle=-90]{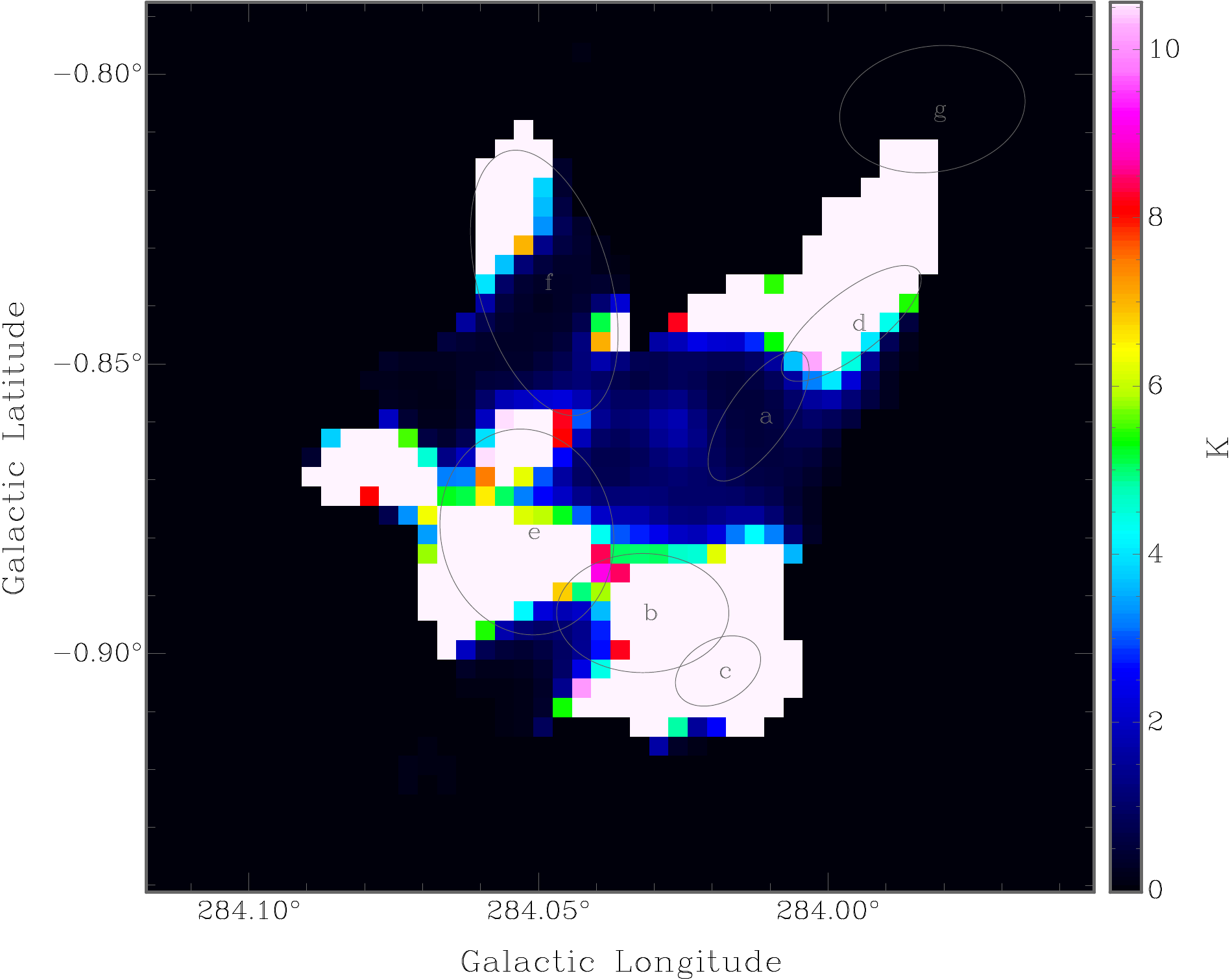}
\label{fig:subfigure8}}
\caption{BYF\,40 maps of $T_{\rm ex}$ and its uncertainty given by \textsc{PySpecKit}, overlaid by the HCO$^{+}$ ellipses from Paper I.}
\label{tex}
\end{figure*}

\bsp	
\label{lastpage}
\end{document}